\documentclass[a4paper,11pt]{article}

\usepackage{jheppub} 

\usepackage[T1]{fontenc} 
\usepackage{bbm}
\usepackage{mycommands}
\usepackage{multirow}
\title{\boldmath Searching for QGP droplets with high-$p_T$ hadrons and heavy flavor}

\author[a]{Weiyao Ke}
\author[a]{Ivan Vitev}

\affiliation[a]{Theoretical Division, Los Alamos National Laboratory, Los Alamos NM 87545, United States}

\emailAdd{weiyaoke@lanl.gov}
\emailAdd{ivitev@lanl.gov}
\abstract{
The search for the smallest quark-gluon plasma (QGP) droplets 
in nature has motivated recent small collisions system  programs at RHIC and LHC. Unambiguous identification of jet quenching due to final-state interactions is key to confirming QGP formation  in these reactions.
We compute the nuclear modification factors $R_{AA}$ and $R_{p(d)A}$ of charged hadrons and heavy flavor mesons in large (Au-Au, Xe-Xe, Pb-Pb) and small  ($d$-Au, $p$-Pb, O-O) colliding systems, respectively.
Our results include the Cronin effect and initial-state parton energy loss in cold nuclear matter. In the final state, hard partons undergo collisional energy loss and branching that was recently derived using  Soft-Collinear-Effective-Theory with Glauber Gluon (SCET$_{\rm G}$).
In large colliding systems, medium-modified QCD evolution of the fragmentation functions dominates the nuclear correction. As the system size decreases, we find that cold nuclear matter effects, collisional energy loss, and QGP-induced radiations can become equally important.  A systematic scan over the medium size and mass/flavor dependence of $R_{AA}$ provides the opportunity to separate these individual contributions and identify QGP signatures in small systems.
Predictions for $R_{AA}^{h}$, $R_{AA}^{D}$, $R_{AA}^{B}$ in O-O collisions at $\sqrt{s}=7$ TeV are presented with and without the formation of a QGP and contrasted with the corresponding $R_{p(d)A}$ calculations.  Upcoming  single-hadron measurements at the LHC will not only test the O-O predictions for both light and heavy flavor production, but will shed light on the possibly very different dynamics of $p$-A and A-A reactions at similar soft particle production multiplicities.
}

\begin{document} 
\maketitle
\flushbottom

\section{Introduction}\label{sec:intro}
\label{sec:intro}
Jet quenching is an unambiguous signature of quark-gluon plasma (QGP) formation in nuclear collisions and has played a central role in its discovery at the Relativistic Heavy Ion Collider (RHIC)~\cite{PhysRevD.33.717,GYULASSY1990432,PhysRevLett.68.1480,Gyulassy:2003mc,Qin:2015srf}. Hard partons created in  QCD scattering processes undergo multiple collisions with the constituents in the hot QGP medium \cite{THOMA1991128,PhysRevD.44.R2625,Mustafa:2003vh,Mustafa:2004dr,Djordjevic:2006tw,Wang:2006qr,Schenke:2008gg,PhysRevLett.100.072301,Liu:2006ug,Neufeld:2011yh,Neufeld:2014yaa}. The interactions further amplify QCD radiation, which was first studied in the soft gluon emission energy loss limit    \cite{BAIER1997265,Zakharov:1996fv,Zakharov:1997uu,Baier:1998kq,Wiedemann:2000za,Gyulassy:2000fs,Gyulassy:2000er,Wang:2002ri,Arnold:2002ja}. The development of effective theories of QCD that describe parton interactions in matter via the exchange of Glauber 
gluons~\cite{Idilbi:2008vm,Ovanesyan:2011xy,Kang:2016ofv,Makris:2019ttx} has enabled calculations of full in-medium splitting functions~\cite{Ovanesyan:2011kn,Fickinger:2013xwa,Kang:2016ofv}. As a result,  
  medium-modified QCD evolution and parton shower calculations ~\cite{Kang:2014xsa,Chien:2015vja,Cao:2017qpx}  have successfully explained the large factor of 5-10 suppression in the nuclear modification factor $R_{AA}$ of hadron production in Au-Au and Pb-Pb collisions at the RHIC and the LHC,
\begin{eqnarray}
R_{AB} = \frac{dN_{AB\rightarrow h}(p_T)}{\langle T_{AB}\rangle d\sigma_{pp\rightarrow h}(p_T)}.
\end{eqnarray}
Here $\langle T_{AB}\rangle$ is the centrality-averaged geometric overlap function of nuclei $A$ and $B$. Such success has also already motivated recent works to reverse-engineer the detailed transport parameters of jets in the QGP~\cite{Bass:2008rv,PhysRevC.90.014909,Andres:2016iys,Xu:2017obm,Xie:2019oxg} and to use the internal structure of jets to understand the microscopic QGP properties \cite{Chien:2015hda,Kang:2016ehg,Tachibana:2018yae,Li:2017wwc}.

Recently, puzzles have emerged in small colliding systems such as $d$-Au and $p$-Pb collisions. Similar ``collective'' behavior in the pattern of soft particle production that is attributed to QGP evolution in large systems has been observed \cite{CMS:2012qk,ALICE:2012eyl,ATLAS:2012cix,PHENIX:2014fnc,STAR:2015kak, PHENIX:2017nae}, suggesting the possibility of final-state effects. However,  clear evidence of jet quenching has not been observed~\cite{STAR:2003oii,PHENIX:2006njd,STAR:2007poe,ATLAS:2016xpn,CMS:2016xef,ALICE:2017svf,ALICE:2021est}. This puzzle has led to intense discussion of the origin of the apparent collectivity \cite{Schlichting:2016sqo,PhysRevLett.123.039901,Schenke:2019pmk,Zhao:2020pty} and the nature of the medium produced in small systems.
Sensitive experiments have been designed to provide further insight into this problem, such as the geometry scan at RHIC using $p$-Au, $d$-Au, and ${}^3$He-Au \cite{PHENIX:2015idk,PHENIX:2018lia,STAR:2019zaf}, and the upcoming O-O and $p$-O collisions program at the LHC \cite{Citron:2018lsq,Brewer:2021kiv}.
Theoretical predictions of jet quenching in O-O collisions already exist \cite{Liu:2021izt,PhysRevC.102.041901,Huss:2020whe,Zakharov:2021uza} for light and heavy-flavor quenching, using various frameworks, including transport equations and energy loss calculations. The QGP effects are modeled by medium-induced gluon radiations using either higher-twist or the BDMPS-Z formula, some studies also included collisional processes.
Nevertheless, what is still missing from the theory side is an analysis that combines QGP and full cold nuclear matter (CNM) effects and a comparison and contrast of results in $p(d)$-A and A-A reactions. This will be important to  better understand the baseline without QGP formation and identify possibly different dynamics in symmetric vs asymmetric small systems. Finally, we use the in-medium QCD evolution formalism to consistently treat final-state parton shower effects for light and heavy flavor and understand the interplay with collisional energy loss \footnote{Medium-modified DGLAP evolution has been applied to understand the suppression of hadrons at the future Electron-Ion Collider~\cite{Li:2020zbk}}.

This paper aims to analyze the nuclear modification factors in various large and small systems. Calculations will be performed for both light hadrons and heavy-flavor mesons effects including both cold nuclear matter effects and hot QGP final-state effects, collisional energy loss and medium-induced radiation corrections to the baseline QCD factorization formalism.
By fixing the jet-medium interaction parameter $g_s$ in large colliding systems, we make predictions for light and heavy-flavor productions in $d$-Au, $p$-Pb, O-O, with and without the assumption of the existence of a QGP. 

The motivation for this comprehensive analysis is threefold.
First, although  initial-state cold nuclear matter effects are overwhelmed by the QGP-induced modifications in large systems, they can be important in small systems and provide an alternative explanation of the structures of modifications without always resorting to the existence of a QGP. 
Second, the radiative correction can be strongly reduced relative to the collisional effect in a QGP of decreasing size, as we will demonstrate in this paper. Therefore, a complete treatment of hot medium effects has to include elastic collisions. Using the distinct mass dependence of radiative and collisional processes, the difference between light and heavy flavor modifications in small systems can provide a handle on the relative contribution of the two.
Third, a notable source of uncertainty that has hindered theoretical analyses in small systems is the decorrelation of charged particle production, which defines centrality classes, and the nuclear overlap function $\langle T_{pA}\rangle$. 
The large model-dependence in $T_{pA}$ makes it hard to interpret the present jet $p$-Pb modification data.
By looking into various collision geometries, especially the O-O program at the LHC, the normalization uncertainty is expected to be significantly reduced.

The paper is organized as follows. In section \ref{sec:Factor} we review the factorization calculation in nuclear collisions and illustrate final-state and initial-state effects that will be considered.
In section \ref{sec:CNM} we describe the dynamical approach for the cold nuclear matter effect.
In section \ref{sec:FS-QGP}, we review in-medium QCD splitting functions obtained in SCET$_{\rm G}$, the modified QCD evolution for in-medium fragmentation function, and collisional energy loss in a thermal QGP. 
The model of QGP medium simulation is introduced in section \ref{sec:med-QGP}.
Results for $R_{AA}$ in large and small systems are presented in section \ref{sec:res}.
 Finally, we present our conclusions in section \ref{sec:summary}. 
 How to constrain the parameters in the hydrodynamic simulation is discussed in appendix A. We comment on the differences between dynamically calculated CNM effects and nPDF parametrization in appendix B.


\section{Factorization approach with initial and final-state effects}
\label{sec:Factor}
\begin{figure}
    \centering
    \includegraphics[width=.5\textwidth]{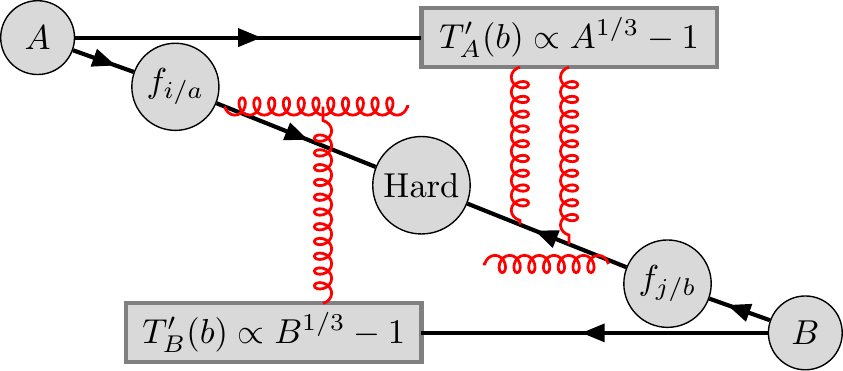}\quad\raisebox{1.5em}{\includegraphics[width=.4\textwidth]{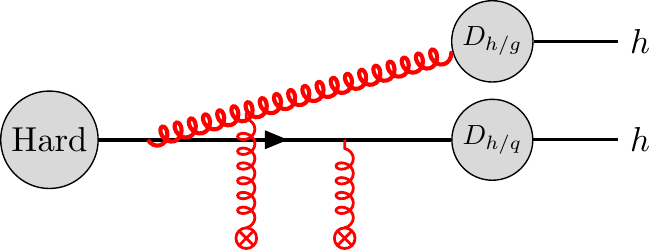}}
    \caption{Left: illustration of cold nuclear matter effects. Vertical gluon lines represent multiple collisions between initial-state partons with the other nucleus. Horizontal lines represent the modified gluon radiation in the initial state. Right: illustration of final-state effects of jets in the quark-gluon plasma.}
    \label{fig:illustration}
\end{figure}

Calculations of hadron and jet production in reactions with nuclei  are based on incorporating medium corrections into the QCD factorization approach. The baseline $p$-$p$ transverse momentum ($p_T$) and rapidity ($y$) differential hadron production cross-section is 
\begin{eqnarray}
\frac{d\sigma_h}{d\bfp^2 dy} &=& \sum_k \int_0^1 \frac{d\sigma_k}{d\bfq^2 dy}(\bfq=\frac{\bfp}{z}, \mu_F, \mu_R) D_{h/k}(z,\mu_F; E) \frac{dz}{z^2}.
\label{eq:FS-formula}
\end{eqnarray}
$D_{h/k}(z,\mu_F)$ is the fragmentation function of parton $k$ into  hadron $h$  carrying momentum fraction $z$. Final-state effects in the QGP modify $D_{h/k}(z,\mu_F)$, and the fragmentation will generally depend on the parton energy $E$ in the rest frame of the medium in addition to the medium transport properties.
$\frac{d\sigma_k}{dq_T^2 dy}$ is the production cross-section of the hard parton and accounts for many-body scattering effects in large nuclei. It can be expressed as  
\begin{eqnarray}
\nonumber
\frac{d\sigma_k}{d\bfq^2 dy} =\frac{4}{s}  \sum_{ij} \int d\eta_{\rm c.m.} &&\int d^2 \bfk_{i}  f_{i/A}(x_i+\Delta x_i, 
 \bfk_{i}; \mu_F)
 \int d^2 \bfk_{j} f_{j/B}(x_j+\Delta x_j, 
 \bfk_j; \mu_F)\\
&& \times \frac{d\sigma_{ij\rightarrow k}}{d\cos\theta_{\rm c.m.}}(x_i x_j s, \cos\theta_{\rm c.m.}; \mu_R).
\label{eq:IS-formula}
\end{eqnarray} 
Equation~\ref{eq:IS-formula} accounts for the fact that the initial parton can acquire a finite transverse momentum $\bfk_i, \bfk_j$ from multiple collisions with the other nucleus, which we treat in the Gaussian approximation~\cite{Baier:1998kq,Qiu:2003pm} that often used to describe the Cronin effect~\cite{Cronin:1974zm}. 
$x_i$ and $x_j$ are the longitudinal momentum that we find from
\begin{eqnarray}
y &=& \frac{1}{2}\ln\frac{x_i}{x_j} + \eta_{\rm c.m.}\\
\left[\bfq_T-\frac{\bfk_i+\bfk_j}{2}\right]^2 &=& \frac{x_ix_js\sin^2\theta_{\rm c.m.}}{4}.
\end{eqnarray}
Another impact of cold nuclear matter is that initial-state partons can lose fractions ($\Delta x_i/x_i$ and $\Delta x_j/x_j$ ) of their energy due to CNM-induced gluon emissions~\cite{Vitev:2007ve}.
The physical picture of equation \ref{eq:IS-formula} and \ref{eq:FS-formula} are illustrated in the left and right panel of figure \ref{fig:illustration}. 
Finally, the coherent scatterings also lead to dynamical shadowing effect that shifts the $x_{i}$ by an amount proportional to the nuclear thickness function~\cite{Qiu:2004da}. It only contributes at small Bjorken-$x$ and low transverse momentum  We will elaborate upon these CNM effects in section \ref{sec:CNM}. 

Finally, $\cos\theta_{\rm c.m.}=\tanh\eta_{\rm c.m.}$ is the polar angle in the center-of-mass frame. The integration has been transformed to the rapidity of the produced particle in the partonic center-of-mass frame $\eta_{\rm c.m.}$.
We will set the factorization and renormalization scale to the transverse momentum of the hard parton $\mu_R = \mu_F = |\bfq| = |\bfp|/z$.

\section{The dynamical approach for cold nuclear matter effects}\label{sec:CNM}
Despite the complicated nature of parton interaction in the nuclear environment, it is possible to model the initial-state effect from QCD interactions.
In reference \cite{Vitev:2007ve}, one considers coherent multiple scatterings (vertical gluons lines in the left panel of figure \ref{fig:illustration}) and induced soft gluon radiation (horizontal gluons lines in the left panel of figure \ref{fig:illustration}) from initial-state partons as they traverse in the nuclear matter before the hard interaction.
The collisions push particle production to slightly larger transverse momentum, resulting in an enhancement at $p_T$ around a few GeV, known as the Cronin effect~\cite{PhysRevD.11.3105}.  
The interaction potential between partons and  cold nuclear matter (CNM) is chosen to be a screened Coulomb potential with typical transverse momentum transfer squared $\mu^2 = 0.12$~GeV$^2$ and the mean free paths $\lambda_g = (C_F/C_A)\lambda_q = 1.5$~fm. As mentioned earlier,  
we employ the Gaussian parametrization of the Cronin effect and account for the power-law tails of the Moliere multiple scattering with a numerical factor $\xi \sim$~few. 
The Cronin effect is sensitive to the shape of the particle spectra and decreases at large colliding energies. For phenomenological applications we also consider 50\%  shorter mean free paths to estimate its uncertainty. At small transverse momenta and small values of Bjorken-$x$ coherent multiple scattering also leads to dynamical shadowing~\cite{Qiu:2003vd,Qiu:2004da}, which we include in the calculation. Coherent power corrections scale as $\Delta x_i/x_i  \sim \mu^2 A^{1/3}/(-u)$  and $ \Delta x_j/x_j \sim \mu^2 B^{1/3}/(-t)$ in A-B reactions and $t$ and $u$ are the partonic Mandelstam variables for the hard scattering process. The Cronin and dynamical shadowing effects disappear at high $p_T$. 

On the contrary, the induced gluon emissions that causes initial-state parton energy loss in CNM continues to be important at high energy~\cite{Vitev:2007ve}
\begin{eqnarray}
x\frac{dN_{\rm IS}}{dxd^2\bfk} &=& \frac{\alpha_s C_R}{\pi^2}\frac{L}{\lambda_g} \int_0^{\frac{\mu p^+}{4}} d^2\bfq\frac{\mu^2}{\pi(\bfq^2+\mu^2)^2} \left[\frac{\bfq^2}{\bfk^2(\bfk-\bfq)^2} -\frac{2(\bfq^2-\bfq\cdot\bfk)}{\bfk^2(\bfk-\bfq)^2} \frac{\sin\frac{\bfk^2 L}{xp^+}}{\frac{\bfk^2L}{xp^+}}\right], \; \; 
\label{eq:CNMeloss}
\end{eqnarray}
as can be seen from its weak dependence on the lightcone momentum $p^+$\footnote{We use the convention $x^\pm = x^0\pm x^3$.}. $x$ is the momentum fraction carried away by the radiated gluon.
$L$ is the path length that parton propagates in the cold nuclear matter before the hard collision.
The same transverse-momentum transfer squared $\mu^2$ and mean free path $\lambda_g$ are used in equation \ref{eq:CNMeloss} as those for the Cronin effect.
The CNM energy loss also causes a shift in the momentum fraction $x$ of the initial parton in equation \ref{eq:IS-formula}
\begin{eqnarray}
\Delta x / x  = \epsilon_{\rm fl} \int_{m_N/p^+}^1 dx \int_{xm_N \leq |\bfk| \leq xp^+} d^2\bfk \; x\frac{dN_{\rm IS}}{dxd^2\bfk} .
\end{eqnarray}
At high energy, the CNM energy loss contribution dominates $\Delta x$ and is proportional to $L$, introducing the centrality dependence. Fluctuations due to multiple gluon emissions reduce the effect of the mean fractional energy loss, which we account for with $\epsilon_{\rm fl} < 1$. For steeply falling final-state spectra  $\epsilon_{\rm fl}$ can be as small as  0.4~\cite{Gyulassy:2001nm}. The energy dependence of the initial parton flux is much more moderate and we use  $\epsilon_{\rm fl} = 0.7$.   For phenomenological applications, we also consider the scenario without initial-state energy loss.

The calculation that includes the Cronin effect, dynamical shadowing, and the CNM energy loss, hereafter referred to as the ``dynamical CNM calculation'' or ``Cronin+$e$loss'',  will be used as the primary model for how the nuclear environment affects the initial parton density $f(x, \bfk)$.
The advantage of the dynamical approach is that one can use only two parameters that control the magnitude of broadening and initial-state energy loss to systematically study the energy and $A$ dependence predicted by QCD. 
In the large $x$ region, the current calculation only implements the isospin effect without parametrizing the anti-shadowing, EMC, and Fermi motion regions.  We expect these effects to be small for the moderate $p_T$ regions of hadron production considered in this paper. 

\begin{figure}
    \centering
    \includegraphics[width=.8\textwidth]{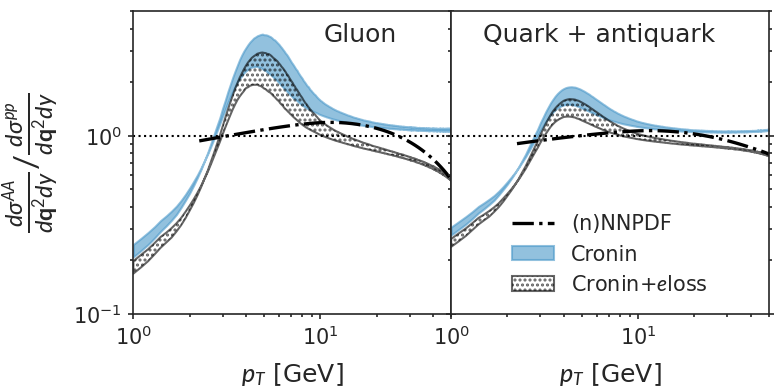}
    \caption{Comparison of cold nuclear matter effects in Au-Au collisions at $\sqrt{s}=200$ GeV from I) (n)NNPDF (black lines); II) dynamical approach without CNM energy loss (blue bands), where bands indicate the variation of the strength of Cronin effect; and III) dynamical approach with CNM energy loss (dotted bands). }
    \label{fig:Raa-cold}
\end{figure}

In figure \ref{fig:Raa-cold} we compute the spectra of partons produced in the hard interaction in Au-Au collisions relative scaled $p$-$p$ baseline at $\sqrt{s}=200$~GeV. The blue bands are dynamical model calculations with Cronin and dynamical shadowing effects only, and shaded bands are results that further include the CNM energy loss. The spread of the bands represents $50\%$ variation in the magnitude of transverse momentum broadening. 
Scattering in nuclear matter introduces a nuclear size-dependent enhancement at moderate $p_T$, while depleting particle production below 2 GeV. 
Dynamical shadowing also contributes to this low-$p_T$ suppression \cite{Qiu:2004da}. 
The CNM energy loss suppresses the spectra at large $p_T$. 
The dynamical CNM calculations are compared to results using collinear nuclear parton distribution functions (nPDF) from the (n)NNPDF Collaboration \cite{Khalek:2022zqe} (black dash-dotted lines). There is no momentum broadening in the collinear nPDF. The small low-$p_T$ suppression comes from the parametrized shadowing effect. At large $p_T$, modifications results from the anti-shadowing, EMC, and the Fermi motion effects included in the nPDF.
Unlike the dynamical CNM model, nNNPDF does not depend on the impact parameter, which is essential to include to study the centrality dependence of $R_{AA}$ and $R_{pA}$ in small colliding systems.
Therefore, we primarily use the dynamical CNM model in this paper. We will compare the impact of using nPDF to the final results in appendix \ref{sec:res:nPDF}.

\section{Final-state QGP  effects}\label{sec:FS-QGP}
\subsection{Medium-modified splitting functions from SCET${}_{\rm G}$}\label{sec:SCETG}
An effective theory of QCD ideally suited to studying jet physics  is Soft-Collinear-Effective-Theory (SCET) \cite{Bauer:2000yr,Beneke:2002ph}. The power counting parameter $\lambda =|{\bf{k}}|/p^+$  
can be thought of as the typical transverse momentum in the jet divided by its large lightcone component. 
In reactions with nuclei, the SCET$_{\rm G}$ theory \cite{Idilbi:2008vm,Ovanesyan:2011xy} was developed to couple the collinear fields to the background nuclear medium via Glauber gluon exchanges. Thus, hadron and jet production, and jet substructure can be described in different strongly-interacting environments without loss of generality.
This framework has been applied to study both the jet broadening~\cite{DEramo:2010wup} and derive the medium-modified QCD splitting functions \cite{Ovanesyan:2011kn,Ovanesyan:2015dop}.

For phenomenological application, the Glauber gluon field is often approximated by a sum of screened color potential over the scattering centers in the medium 
\begin{eqnarray}
V_R^{a}(\bfq) = \sum_i g_s^2\frac{T^a(R)\otimes T^a(i)}{\bfq^2+m_D^2} e^{-i\bfq\cdot \bfz-i \frac{1}{2} q^-z^+}.
\end{eqnarray}
Here, $R$ denotes the color representation of the collinear parton while $i$ is the representation of the color charge of the medium quasi-particle. 
$g_s$ is the jet-medium coupling constant. $m_D^2 = g_s^2 (1+\frac{N_f}{6}) T^2$ is the Deybe screening mass of the plasma. There  is no momentum exchange in the
$p^+$ direction that scales as $\lambda$ or stronger. 
The phase factor contains the position ($\bfz, z^+$)information of the medium color charge.
The differential collision rate between a collinear quark (or a gluon) and the QGP is
\begin{eqnarray}
\frac{dR_{q,g}(x)}{d^2\bfq} =  \frac{\lambda_{q,g}^{-1}}{\pi}\frac{m_D^2}{(\bfq^2+m_D^2)^2},
\end{eqnarray}
 with $\lambda_{g}^{-1} = \frac{C_F}{C_A}\lambda_{q}^{-1} = \frac{\zeta(3)T^3}{\pi^2} \left(2 d_A + \frac{3}{4} \frac{C_F}{C_A} 4 d_F N_f  \right)\sigma_{gg}$ being the inverse mean-free-path of the parton in the plasma and $\sigma_{gg} = \frac{9}{32\pi}\frac{g_s^4 }{m_D^2}$ is the screened Columb cross-section of gluons.
$\zeta(3)\approx 1.2$, $d_A=N_c^2-1, d_F=N_c$, and we choose an effective $N_f=2$ for the QGP.

The modified QCD splitting functions have been obtained to first order in opacity explicitly \cite{Ovanesyan:2011xy,Ovanesyan:2011kn} and an iterative approach has been developed to generalize them to higher opacity orders \cite{Sievert:2019cwq}. For light partons, we take the full splitting function to first order, which we quote for completeness from Ref. \cite{Ovanesyan:2011kn}. The double differential spectrum of the quark to quark+gluon branching is
\begin{eqnarray}
\frac{dN^{\rm med}_{qq}}{dx d\bfk^2} &\equiv& P_{qq}(x, \bfk^2)\int_{0}^{\infty} d \Delta z \int d^2\bfq \frac{dR_g(\Delta z)}{d^2\bfq}
\nonumber\\
&&\left\{\left[\iB\cdot\left(\iB-\iC\right)+\frac{1}{N_c^2}\iB\cdot\left(\iA-\iB\right)\right][1-\cos(\omega_1 \Delta z)]\right.\nonumber\\
&&+ \iC\cdot\left(2\iC-\iA-\iB\right)[1-\cos(\omega_2 \Delta z)] + \iB\cdot\iC[1-\cos(\omega_3 \Delta z)]\nonumber\\
&& \left.-\iA\cdot \left(\iA-\iD\right)[1-\cos(\omega_4 \Delta z)] - \iA\cdot\iD[1-\cos(\omega_5\Delta z)]\right\},
\label{eq:Pij_med_1}
\end{eqnarray}
where the gluon carries momentum fraction $x$. Note that this differs from the standard high energy notation and is done to make contact with the much studies energy loss soft gluon emission limit when $x\rightarrow 0$. The quark to gluon+quark splitting function is obtained by $\frac{dN^{\rm med}_{gq}}{dx d\bfk^2} (x) = \frac{dN^{\rm med}_{qq}}{dx d\bfk^2}(1-x)$.
For gluon to quark+quark and gluon to gluon+gluon splittings, the distributions are
\begin{eqnarray}
\frac{dN^{\rm med}_{\{gg, qg\}}}{dx d\bfk^2} &\equiv& P_{\{gg,qg\}}(x, \bfk^2)\int_{0}^{\infty} d \Delta z \int d^2\bfq \frac{dR_{\{g,q\}}(\Delta z)}{d^2\bfq}
\nonumber\\
&&\left\{2\left[\iB\cdot\left(\iB-\iA\right)+\left\{-\frac{1}{2}, \frac{1}{d_A}\right\}\iB\cdot\left(\iC-\iA\right)\right][1-\cos(\omega_1 \Delta z)]\right.\nonumber\\
&&+ 2\iC\cdot\left(\iC+\iB-2\iA\right)[1-\cos(\omega_2 \Delta z)] - 2\iB\cdot\iC[1-\cos(\omega_3 \Delta z)]\nonumber\\
&& \left.+2\iA\cdot \left(\iA-\iD\right)[1-\cos(\omega_4 \Delta z)] +2 \iA\cdot\iD[1-\cos(\omega_5 \Delta z]\right\}.
\label{eq:Pij_med_2}
\end{eqnarray}
In the above expressions, we have defined vectors
\begin{eqnarray}
\mathbf{A} =\bfk,~\mathbf{B} = \bfk+x\bfq,~\mathbf{C} = \bfk-(1-x)\bfq,~\mathbf{D} = \bfk-\bfq,
\end{eqnarray}
and frequencies (which are inverse formation times)
\begin{eqnarray}
\omega_1 = \frac{\mathbf{B}^2}{x(1-x)p^+}, &&\omega_2 = \frac{\mathbf{C}^2}{x(1-x)p^+},\nonumber\\
\omega_3 = \frac{\mathbf{C}^2-\mathbf{B}^2}{x(1-x)p^+}, && \omega_4 = \frac{\mathbf{A}^2}{x(1-x)p^+},~ \omega_5 = \frac{\mathbf{A}^2-\mathbf{D}^2}{x(1-x)p^+}.
\end{eqnarray}
$P_{qq}, P_{gq}, P_{gg}, P_{qg}$ are the standard splitting functions in the vacuum at leading order,
\begin{eqnarray}
P_{qq} = \frac{\alpha_s(\mu^2)}{2\pi} C_F \frac{1+(1-x)^2}{x},&& P_{gg} = \frac{\alpha_s(\mu^2)}{2\pi} C_A \frac{1+x^4+(1-x)^4}{x(1-x)}, \nonumber \\
P_{gq} = \frac{\alpha_s(\mu^2)}{2\pi} C_F \frac{1+x^2}{1-x}, && P_{qg} = \frac{\alpha_s(\mu^2)}{2\pi} T_F [x^2+(1-x)^2].
\label{eq:vacPijinmed}
\end{eqnarray}
The coupling constants associated with the vacuum splitting functions are evaluated at $\mu^2=\bfk^2$ in this study, while the jet-medium coupling parameter $g_s$ that goes into the collisions rates $dR/d\bfq^2$ is taken as a free parameter.
The $\Delta z$ integration in equations \ref{eq:Pij_med_1} and \ref{eq:Pij_med_2} starts from the hard production time and is weighted by the collision rates along the trajectory of the parton. The space-time temperature profile of the QGP is obtained in a hydrodynamic simulation (see section \ref{sec:med-QGP}). 
Equations \ref{eq:Pij_med_1} and \ref{eq:Pij_med_2} are the medium corrections to the real emission function of the splitting. In section \ref{sec:FS-QGP:DGLAP}, we will discuss the inclusion of both real emission and virtual corrections in the DGLAP evolution equation.

The vacuum splitting functions involving heavy quark $H$ ($H=c,b$) with mass $M$ are
\begin{eqnarray}
P_{HH} &=& \frac{\alpha_s(\mu^2)}{2\pi} C_F  \left[ \frac{1+(1-x)^2}{x} - \frac{2x(1-x)M^2}{\bfk^2+x^2M^2}\right], \label{eq:vacPijvacHQ-1} \\
P_{gH} &=& \frac{\alpha_s(\mu^2)}{2\pi} C_F  \left[\frac{1+x^2}{1-x} - \frac{2x(1-x)M^2}{\bfk^2+(1-x)^2M^2}\right], \label{eq:vacPijvacHQ-2}\\
P_{Hg} &=& \frac{\alpha_s(\mu^2)}{2\pi} T_F \left[x^2+(1-x)^2 + \frac{2x(1-x)M^2}{\bfk^2+M^2}\right].  \label{eq:vacPijvacHQ-3}
\end{eqnarray}
The $x$-dependence can be significantly modified compared to light flavors when $\bfk^2\lesssim M^2$.
The formulas for the medium corrections to equations \ref{eq:vacPijvacHQ-1}-\ref{eq:vacPijvacHQ-3} have been derived in reference \cite{Kang:2016ofv}. Though we will not write down the full in-medium expressions, we emphasize that the heavy quark mass not only introduces corrections to the propagators and the interference phases but also generates many new terms proportional to the mass.

\begin{figure}
    \centering
    \includegraphics[width=1\textwidth]{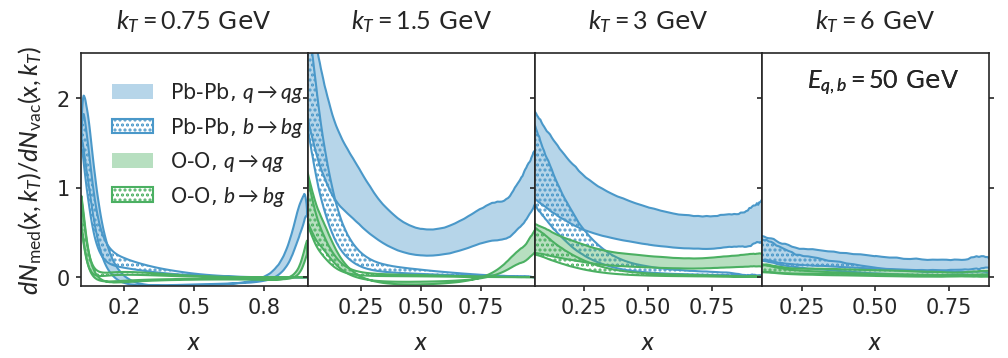}
    \caption{The medium modification to the QCD splitting function $\frac{dN_{\rm med}}{dx d^2\bfk}$ divided by $\frac{dN_{\rm vac}}{dx d^2\bfk}$ for the  $q\rightarrow q+g$ (filled bands) and $b\rightarrow b+g$ (dotted bands) channels. The corrections are shown as a function of gluon momentum fraction $x$ at four different $k_T$ for each panel.
    The bands show the variation of the correction within $1.6<g_s<2.0$. Blue and green colors label the results for central Pb-Pb collisions at 5.02 TeV and central O-O collisions at 7 TeV, respectively.}
    \label{fig:Pij}
\end{figure}

We demonstrate the impact of mass corrections in the medium with numerical results, as shown in figure~\ref{fig:Pij}.
In figure~\ref{fig:Pij}, we present the medium-correction to $q\rightarrow q+g$ splitting functions in 0-5\% central Pb-Pb collisions at 5.02 TeV and in 0-5\% O-O at 7 TeV.
We have factored out the the vacuum-like $P_{qq}$ and $P_{bb}$ kernels given by equation~\ref{eq:vacPijinmed} and equation~\ref{eq:vacPijvacHQ-1}. Medium corrections to light parton branching are suppressed by the Landau-Pomeranchuk-Migdal interference effect at large $x$. For the bottom quark, the vacuum splitting functions $P_{bb}$ further shows the so-called dead-cone effect that suppress radiations in the phase-space region $x>|\bfk|/M$ relative to $P_{qq}$. However, after factoring out the vacuum factor, the ratio still displays strong mass modifications~\cite{Li:2017wwc}. Only ratios in the energy loss region $x\rightarrow 0$ are comparable to that of the light quark.

Looking at the $|\bfk|$-dependence in figure \ref{fig:Pij} or simply at equations \ref{eq:Pij_med_1}, medium-induced branchings are suppressed at large $\bfk$ by at least another power of $1/\bfk^2$ as compared to the vacuum radiation. So, in principle, they do not contribute additional $\ln Q^2$ enhancement upon integration for asymptotic energies.
Its contribution peaks when $|\bfk|$ are comparable to the typical size of $|\bfq|$. In central Pb-Pb collisions, the correction at its peak can be much larger than the vacuum contribution. In central O-O collisions at 7 TeV, it is estimated to be about a factor of three smaller than that in central Pb-Pb at 5 TeV, assuming QGP effects do exist in O-O.

\subsection{Collisional energy loss}\label{sec:el}
In obtaining equations \ref{eq:Pij_med_1} and \ref{eq:Pij_med_2}, one assumes that $p^+$ is conserved. This is a good approximation for the computation of radiative correction because a mismatch in $p^+$ due to collisional energy losses is expected to be subleading in powers of $\lambda$.
However, collisional energy loss should be taken into account to compute $R_{AA}$ at intermediate and small $p_T$~\cite{Djordjevic:2006tw,PhysRevD.77.114017,Neufeld:2011yh}. It was found that it can be comparable to the radiative energy loss for partons up to $p_T = 10$ to $20$ GeV/$c$. For heavy flavor particles, due to the reduced phase space for radiation when $p_T$ is only a few times the heavy quark mass $M$,  the dead cone effect renders collisional energy loss even more important to describe heavy meson suppression at the intermediate $p_T$. Finally, as we will see immediately, the different path length dependence of induced radiation and collisions energy dissipation makes the latter an indispensable component in the analysis of small collision systems.

We take the collisional energy loss obtained in hard-thermal loop calculations \cite{PhysRevD.44.R2625}. The energy loss of a quark per unit length in a weakly-coupled thermal plasma is given by,
\begin{eqnarray}
\frac{dE_{\rm el}}{d \Delta z } = \frac{C_F}{4}\left(1+\frac{N_f}{6}\right) \alpha_s(ET)g_s^2 T^2 \ln\left(\frac{ET}{m_D^2}\right)\left( \frac{1}{v} - \frac{1-v^2}{2v^2}\ln\frac{1+v}{1-v}\right) , 
\end{eqnarray}
with $v=p/E$ being the velocity of the parton in the rest frame of the QGP, applied to both heavy and light quarks. Gluon energy loss is related to that of the quark via the $C_A/C_F$ quadratic Casimir ratio.
The running coupling value at one loop is used in the above expression \cite{PhysRevD.77.114017}, where the maximum $\alpha_s$ is cut-off at $g_s^2/(4\pi)$.
We remark that the running coupling effect cancels the $\ln(ET/m_D^2)$ enhancement from phase-space integration and results in an approximately energy-independent collisional energy loss.

In this study, we use averaged collisional energy loss $\Delta E = \int_{x_{\perp,0}}^{x_{\perp,0}+\Delta z_{\rm max}\hat{\phi}} \frac{dE_{\rm coll}}{d \Delta z} d \Delta z$ obtained by averaging over the production location $x_{\perp,0}$ and orientation $\hat{\phi}$ of a parton with initial energy $E$.
We will apply the averaged energy loss to the partons created in the hard process before invoking the in-medium splitting function modification of the fragmentation function.
The justifications of this approximation are that 1) collisional energy loss is almost $E$-independent, and 2) a $\Delta E_{\rm el}$ mismatch will only cause a small difference in the QCD evolution so long as $\Delta E_{\rm el}\ll Q=|\bfq|$.

\subsection{Medium size dependence of radiative and collisional energy loss}\label{sec:FS:el-rad}
While we employs the full expressions (equations  \ref{eq:Pij_med_1} and \ref{eq:Pij_med_2}) in the calculation, it is instructive to look at the radiative parton energy loss obtained in the $x\rightarrow 0$ limit. The energy loss fraction $\Delta E/E = \int x \frac{dN_{\rm med}}{dx} dx$ in this approximation is
\begin{eqnarray}
\frac{\Delta E_{\rm rad}}{\alpha_s C_R E} 
&=& \int d\Delta z \frac{2\mu^2 }{E} \frac{\Delta z}{\lambda_g} \int \frac{du_t}{u_t^2} \left[\gamma_E+\ln(u_t) +\frac{\pi}{2}\sin(u_t) - \cos(u_t)\textrm{Ci}(u_t)
 - \sin(u_t)\textrm{Si}(u_t)\right] \nonumber \\
 &\sim& \int d\Delta z \frac{2\mu^2}{E} \frac{\Delta z}{\lambda_g}  \ln \frac{2E}{\mu^2 \Delta z}  ,
\end{eqnarray}
with $u_t = \frac{\mu(\Delta z)^2 \Delta z}{2xE}$ and the second line for the asymptotic behavior at high energy. Compared to the scaling of the elastic energy loss fraction
\begin{eqnarray}
\frac{\Delta E_{\rm el}}{\alpha_s C_R E} \propto \frac{\int \mu^2 d\Delta z}{E},
\end{eqnarray}
one sees that radiative energy loss only dominates over the elastic one by $\alpha_s \ln E$. Furthermore, they scale differently with medium size. 
For example, in a QGP that undergoes Bjorken expansion such that $T^3 \tau = T_0^3 \tau_0$ and $\Delta z = \tau-\tau_0$, the typical momentum transfer and the mean free path evolve with proper time as $\mu^2 = \mu^2_0 (\tau_0/\tau)^{2/3}$ and $\lambda_g = \lambda_{g,0} (\tau/\tau_0)^{1/3}$. Therefore, the radiative energy loss fraction
\begin{eqnarray}
\frac{\Delta E_{\rm rad}}{E} \propto  \int_{\tau_0}^{\tau_0+L} \frac{\mu^2}{\lambda_g} \Delta z d \Delta z \propto L
\end{eqnarray}
only scales linearly with size; while the elastic energy loss fraction 
\begin{eqnarray}
\frac{\Delta E_{\rm el}}{E} \propto \int_{\tau_0}^{\tau_0+L}  \mu^2 d \Delta z \propto L^{1/3},
\end{eqnarray}
changes much slower with $L$.
 Consequently, one expects that collisional processes becomes increasingly important in a small-sized QGP. 

\subsection{Modified QCD evolution equations for in-medium fragmentation}\label{sec:FS-QGP:DGLAP}
To compute hadron production in a nuclear environment, we take the modified DGLAP approach \cite{Wang:2009qb} to evolve the vacuum fragmentation function from an initial non-perturbative scale $Q_0$ to $Q=p_T + \Delta E_{\rm el}$ with medium modified QCD splitting functions described in section \ref{sec:SCETG}. 

\paragraph{Evolution in the vacuum.} The evolution equation for the fragmentation function $D_{h/i}^0$ of hadron specie $h$ from parton $i$ produced in the vacuum is
\begin{eqnarray}
\frac{\partial D_{h/i}^0(z, Q^2)}{\partial \ln Q^2} = \sum_j \int_z^{1} \frac{dx}{x} \left[P_{ji}'(x\rightarrow 1-x, Q^2) + d_{ji}(Q^2)\delta(1-x) \right]D_{h/j}\left(\frac{z}{x}, Q^2\right). \;
\label{eq:DGLAP}
\end{eqnarray}
$i=q, g$ and $H$, with $H$ denotes heavy quarks.
Note that we adhere to the standard high energy definition of the momentum fraction $x$ in the DGLAP equation:
$z$ is the momentum fraction of the produced hadron relative to the momentum of parton $i$, $x$ is the momentum fraction retained by the parent parton $i$. 
$P_{ji}'$ are QCD splitting functions as defined in equations \ref{eq:vacPijinmed} but with $x \rightarrow 1-x$ as indicated, and correspondingly singularities that go as $\frac{1}{1-x}$ are replaced by the plus function $\frac{1}{(1-x)_+}$.
The running coupling is evaluated at $\mu^2=Q^2$ for the vacuum evolution.
The $d_{ji}\delta(1-x)$ terms are virtual corrections that only appear in the diagonal terms ($d_{gq}=d_{gH}=d_{qg}=d_{Hg}=0$). The diagonal terms $d_{qq}$, $d_{HH}$ and $d_{gg}$ can be determined by imposing the conservation of flavor and lightcone momentum \cite{Kang:2014xsa},
\begin{eqnarray}
0&=&\int_0^1\left[ P_{qq}'(1-x, Q^2) + d_{qq}\delta(1-x)\right] dx,\\
0&=&\int_0^1 \left[ P_{HH}'(1-x, Q^2) + d_{HH}\delta(1-x)\right]  dx,\\
0&=&\int_0^1 x \left[ P_{gg}'(1-x, Q^2) + \sum_{i=u,d,s,c,b}P_{ig}'(1-x, Q^2)  + d_{gg}\delta(1-x)\right] dx.
\end{eqnarray}
The above equations solve to
\begin{eqnarray}
d_{qq}(Q^2) &=& \frac{\alpha_s(Q^2)}{2\pi} C_F\frac{3}{2},  \\
d_{HH}(Q^2, r) &=& \frac{\alpha_s(Q^2)}{2\pi} C_F c_{HH}(r),\\
d_{gg}(Q^2, r) &=& \frac{\alpha_s(Q^2)}{2\pi} \left[\frac{11}{6}N_c - N_f T_F\frac{2}{3} + \sum_{H=c,b}T_F c_{gH}(r)\right].
\end{eqnarray}
$d_{gg}$ and $d_{HH}$ also depends on the ratio $r = M/Q$, if allowed by kinematics, with
\begin{eqnarray}
c_{gH}(r) &=& F\left(\frac{1+\sqrt{1-4r^2}}{2}\right) - F\left(\frac{1-\sqrt{1-4r^2}}{2}\right) - 2r^2 \sqrt{1-4r^2},\\
F(x) &=& -x^4 + \frac{4}{3}x^3 - x^2,\\
c_{HH}(r) &=& \frac{1}{1+r^2} + \frac{2r^2+1}{2(1+r^2)^2}  +  \frac{2r^2}{1+r^2} - 2\ln\frac{1}{1+r^2}.
\end{eqnarray}
Finally, the relation between $Q^2$ and $x, \bfk$ and the allowed kinematic ranges are summarized in table \ref{tab:Q2_and_kt2} for each channel.
\begin{table}[h!]
\renewcommand{\arraystretch}{1.5}
    \centering
    \begin{tabular}{c|c|c|c}
    \hline
    Channel & $Q^2$ definition & $Q^2$ constraints & $x$ constraints\\
    \hline
    all light flavors & $\frac{\bfk^2}{x(1-x)}$ & $Q^2>\LambdaQCD^2$ & -\\
    $HH$     & $\frac{\bfk^2+(1-x)^2M^2}{x(1-x)}$ & $Q^2>\frac{1-x}{x}M^2$ & $x > \frac{r^2}{1+r^2}$ \\
    $gH$     & $\frac{\bfk^2+x^2 M^2}{x(1-x)}$  & $Q^2>\frac{x}{1-x}M^2$ & $x < \frac{1}{1+r^2}$\\    
    \multirow{2}{*}{$Hg$}    & \multirow{2}{*}{$\frac{\bfk^2+M^2}{x(1-x)}$}  & \multirow{2}{*}{$Q^2>\frac{M^2}{x(1-x)}\geq 4M^2$} & 
    $x>\frac{1-\sqrt{1-4r^2}}{2}$\\
    & & & $x<\frac{1+\sqrt{1-4r^2}}{2}$\\
    \hline
    \end{tabular}
    \caption{$Q^2$ in terms of splitting kinematics and the kinematic ranges of $Q^2$ (or of $x$).}
    \label{tab:Q2_and_kt2}
\end{table}

\paragraph{Evolution in the medium.} For the QCD evolution in the medium, the splitting functions and virtual corrections in equation \ref{eq:DGLAP} are replaced by the medium-modified ones,
\begin{eqnarray}
P_{ji}' &\rightarrow& P_{ji}' + \bfk^2\frac{{dN_{ji}^{\rm med}}'}{dx d\bfk^2}\quad\textrm{with~}x\rightarrow 1-x,
\end{eqnarray}
and
\begin{eqnarray}
d_{ji}(Q^2)&\rightarrow& d_{ji}(Q^2) + d_{ji}^{\rm med}(Q^2).
\end{eqnarray}
${dN_{ji}^{\rm med}}'$ can be obtained from equations \ref{eq:Pij_med_1} and \ref{eq:Pij_med_2} with the substitution $x\rightarrow 1-x$. The $\frac{1}{1-x}$ terms are then factored out where applicable and supplemented by the ``plus''-function prescription. 
$d_{qq}^{\rm med}, d_{HH}^{\rm med}$ and $d_{gg}^{\rm med}$ are similarly obtained by imposing flavor and lightcone-momentum conservation for the medium corrections,
\begin{eqnarray}
0&=&\int_0^1\left[ \bfk^2\frac{{dN_{qq}^{\rm med}}'}{dx d\bfk^2}+ d_{qq}^{\rm med}\delta(1-x) \right] dx,\\
0&=&\int_0^1\left[ \bfk^2\frac{{dN_{HH}^{\rm med}}'}{dx d\bfk^2}+ d_{HH}^{\rm med}\delta(1-x) \right] dx,\\
0&=&\int_0^1x \left[ \bfk^2\frac{{dN_{gg}^{\rm med}}'}{dx d\bfk^2} + \sum_{i=u,d,s,c,b} \bfk^2\frac{{dN_{ig}^{\rm med}}'}{dx d\bfk^2}+ d_{gg}^{\rm med}\delta(1-x) \right] dx.
\end{eqnarray}

The medium-induced correction to the splitting function does not introduce a  $\ln Q^2$ dependence as large as the vacuum one because it decays as $1/\bfk^4$ or faster at large $\bfk^2$ and is screened by the Debye mass $m_D$ at small $\bfk^2$. In the low-$Q^2$ region, the medium modifications become comparable to or can even dominate over the vacuum contribution.
When this happens, one has to consider how to implement such corrections. For example, in \cite{Chang:2014fba}, the authors only applies the modified DGLAP equation to the region $Q>1$ GeV, and use the medium-modified QCD splitting function below 1 GeV to build an in-medium initial condition of the evolution. Other methods treat the low-$Q^2$ region in a transport approach \cite{Blaizot:2013vha,Cao:2016gvr,Putschke:2019yrg,Ke:2020clc} where multiple medium-induced emissions are generated sequentially in a time-ordered fashion because the number of soft emissions is enhanced by the medium size.
In our calculation, we notice that with the choice $Q^2 = \bfk^2/(x(1-x))$, $Q^2$ is inversely proportional to the formation time of the branching $\tau_f = p^+/Q^2$. For soft emissions that do not significantly change $p^+$, the $Q^2$-ordered evolution is the same as a formation-time ordered approach to compute radiative parton energy loss. However, they are not the same for energetic splittings that take a large fraction of the parton energy.
We further remark that the QCD evolution approach can be applied to regions where the branching fraction $x$ is large and regions where the medium-correction is negative due to interference, which are beyond the scope of the transport equation.
This may be important for small collisions systems, as the interference effects are very sensitive to a small path length. 
Therefore, we consider the QCD evolution approach to be a better choice for the system-size scan down to small collisions systems such as O-O and $p$-Pb.

\begin{figure}
    \centering
    \includegraphics[width=.8\textwidth]{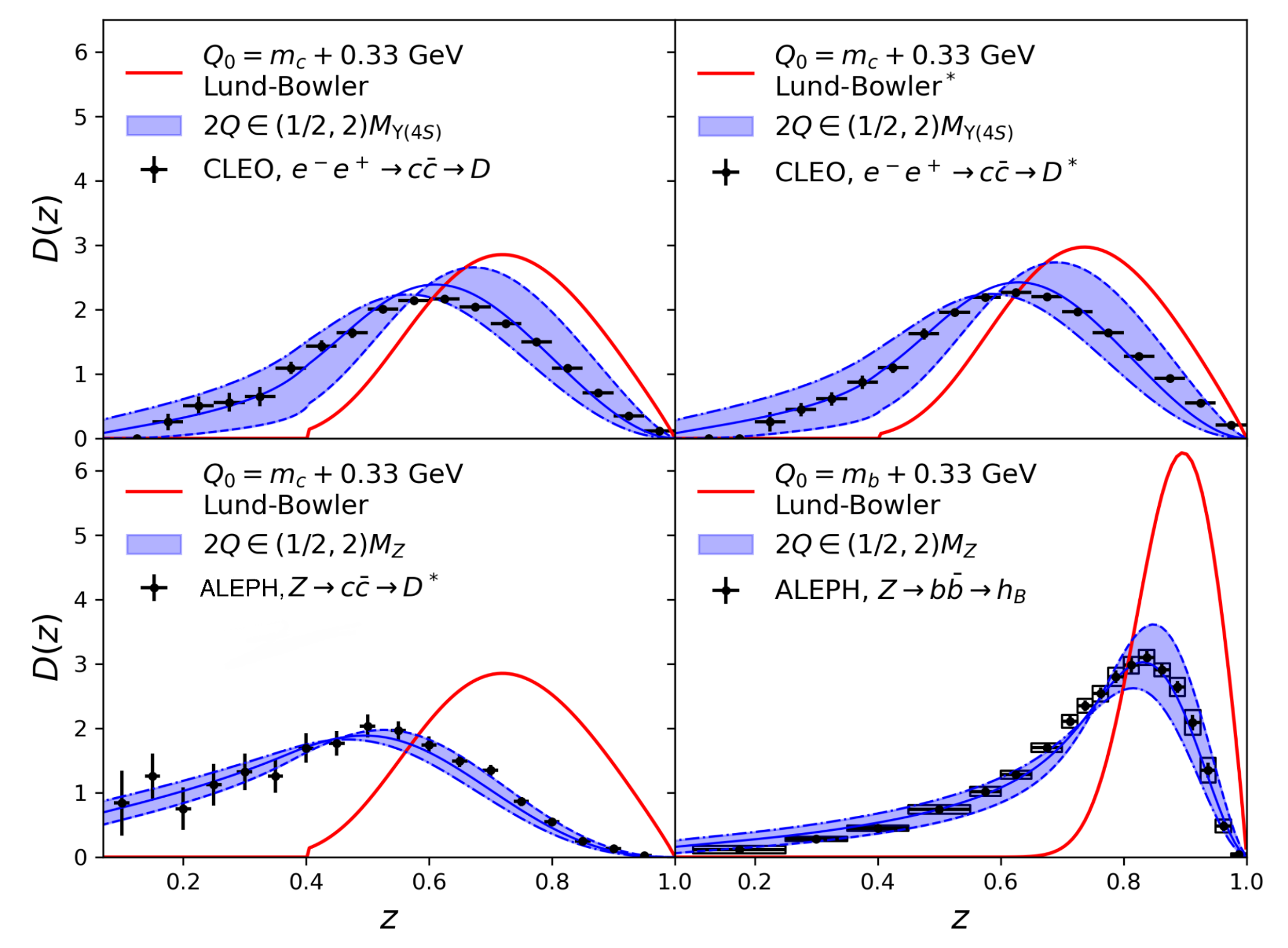}[h!]
    \caption{Heavy-meson fragmentation function evolved from the Lund-Bowler-type initial condition from $Q_0=M_H+m_q$ to $Q$ are compared to data. CLEO~\cite{CLEO:2004enr} data on $D$ and $D^*$ fragmentation are obtained at the $\Upsilon(4S)$ threshold of about $Q=10$ GeV. ALEPH~\cite{ALEPH:1999syy, ALEPH:2001pfo} experiments measure the charm and bottom fragmentation function to mesons at $M_z\approx 92$ GeV. The bands correspond to variation of $Q$ by factors of $1/2$ and $2$. Red lines are the Lund-Bowerler initial condition.}
    \label{fig:HF-FF}
\end{figure}
\begin{figure}
    \centering
    \includegraphics[width=.8\textwidth]{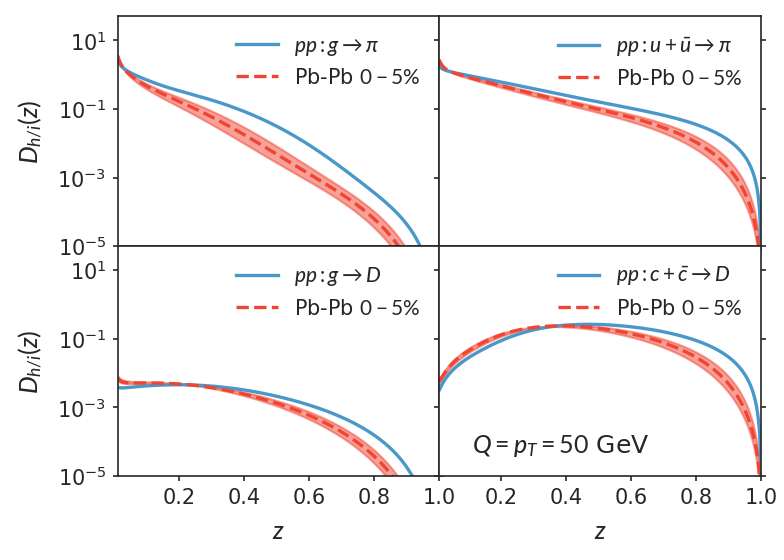}[h!]
    \caption{Modified fragmentation functions in Pb-Pb 0-5\% at 5.02 TeV. From left to right, top to bottom panels are for the $g\rightarrow \pi$, $u+\bar{u}\rightarrow \pi$, $g\rightarrow D$, and $c+\bar{c}\rightarrow D$ channels, respectively. The blue solid lines are the evolved ($Q=p_T$) fragmentation functions in the vacuum for proton-proton collisions. Red dashed lines with bands are results in Pb-Pb collisions, varying $g_s = 1.8 \pm 0.2$. }
    \label{fig:frag}
\end{figure}

Both the vacuum and in-medium evolution use vacuum fragmentation functions at $Q_0=0.4$ GeV as the initial condition. The QCD evolution approach, similar to the traditional energy loss approach, assumes that hadronization happens outside of the medium. 
However, for heavy mesons, the formation time can be significantly shortened by the large mass~\cite{Adil:2006ra}. This may have additional phenomenally consequences and we will come back to this point in section~\ref{sec:res:large}.
We take the charged pion fragmentation functions as parametrized in reference \cite{PhysRevD.91.014035} at $Q=1$ GeV. They are evolved backward from $Q=1$ GeV to $Q_0=0.4$ GeV to provide the common initial condition for evolution in the vacuum and medium.
For heavy mesons, we use the Lund-Bowler function~\cite{Bowler:1981sb} as the initial condition,
\begin{eqnarray}
D(z) = z^{-1-b M_\perp^2} (1-z)^a e^{-\frac{b M_\perp^2}{z}},
\end{eqnarray}
with parameters $a=0.68, b=0.98$ taken from Pythia8 \cite{Sjostrand:2014zea} and $M_\perp^2 \approx M_h^2 + (0.7~{\rm GeV})^2$.
We have verified that the evolved heavy meson fragmentation functions with Lund-Bowler type initial conditions provide a reasonable description to heavy-meson fragmentation measurements by CLEO \cite{CLEO:2004enr} Collaboration at $Q=M_{\Upsilon(4S)}/2$ and ALEPH \cite{ALEPH:1999syy, ALEPH:2001pfo} Collaboration at $Q=M_Z/2$. In figure \ref{fig:HF-FF}, the red lines are the initial conditions and the blue lines are results evolved to $Q=p_T=M_z/2$ for ALEPH experiment and $Q=p_T=M_{\Upsilon(4S)}/2$ for CLEO experiment. The blue bands denote variation $p_T/2<Q<2p_T$.

In figure \ref{fig:frag}, we compare the  fragmentation functions in $p$-$p$ (blue solid lines) and in 0-5\% central Pb-Pb collisions (red dashed lines and bands) for four channels evolved from $Q=Q_0$ to $Q=p_T=50$ GeV. 
The blue solid lines are the evolved results in the vacuum, and the red dashed lines with bands are results evolved in the medium with $g_s=1.6, 1.8, 2.0$.
Compared to $D(z)$ in the vacuum, medium-modified DGLAP evolution ``red-shifts'' $D(z)$. 
In the calculation of inclusive hadron spectra, $D(z)$ is always folded with a steep falling partonic cross-section $d\sigma/d\bfq \sim 1/\bfq^n, n\gg1$. 
The resulting spectra depend on the integral $\int_0^1 z^{n-1} D(z) dz$. 
Because $n\gg 1$, $R_{AA}$ of both light and heavy mesons are mostly sensitive to the modification in the region $z\lesssim 1$, even though the pion fragmentation functions (top row) are much softer than those of heavy mesons (bottom row).

\section{Dynamical simulations of QGP and its existence in small systems}\label{sec:med-QGP}
Finally, we discuss the model for the medium evolution.
The dynamical simulation of the quark-gluon plasma produced in nuclear collisions is performed using the Duke hic-eventgen code package \cite{Bernhard:2018hnz}.
In this calculation, the TRENTo initial condition model of the collision geometry provides the energy deposition profiles at the proper time $\tau=0^+$.
The model of quark-gluon plasma dynamics consists of a pre-equilibrium  stage modeled by free-streaming \cite{Broniowski:2008qk}, followed by the 2+1D boost-invariant relativistic viscous hydrodynamics\cite{Song:2007ux,SHEN201661}. Finally, the hydrodynamic fields are particlized into hadrons at transition temperate $T_{\rm sw}$ slightly below the pseudo-critical temperature of the QGP equation of state \cite{Bazavov:2014pvz}, and the hadronic interactions are handled by the Ultra-relativistic-Quantum-Molecular-Dynamics (UrQMD) \cite{Bass:1998ca,Bleicher:1999xi}.
The model parameters have been tuned to the experimental measurement of particle production, flows, and correlations in previous studies~\cite{Bernhard:2018hnz}.

In the current study, it will be very computationally intensive to obtain the full splitting functions and perform DGLAP evolution on an event-by-event basis. Therefore, we simulate events with centrality-averaged initial conditions. For this reason, the scale parameter (normalization) of the TRENTo energy deposition model is re-tuned for each system to reproduce the centrality-dependent charged particle yield and transverse energy. 
Because no data is available for O-O collisions at 7 TeV, we interpolate the normalization tuned at RHIC and LHC energies using a third-degree polynomial in $\ln\sqrt{s}$ to predict the normalization at $7$ TeV. The details can be found in appendix \ref{sec:app:hydro}. 

Now, we make an important remark on the QGP effects in small systems.
This may naively seem to be an unnecessary discussion because the hydrodynamic simulations already provide the time evolution of the temperature profile. One can, in principle, use this to compare the medium temperature and the QGP pseudo-critical temperature $T_c$ to determine if the jet propagates in the QGP phase or in a hadronic phase. In fact, for those high-multiplicity events in small system collisions, the simulated medium temperature starts from a point well above $T_c$.
However, the definition of temperature in the hydrodynamics-based simulation bears certain ambiguity when the system is far from equilibrium, for example, in small colliding systems.
In the hydrodynamic picture, energy density is converted into temperature using the lattice QCD EoS, which means that the number of scattering centers defined in this manner would approach the thermal limit. If the system is far from equilibrium, the density of scattering centers can significantly deviate from this expectation.
In this study, we will therefore investigate two extreme limits of small colliding systems.
\begin{itemize}
\itemsep0.3em
    \item Calculations with cold nuclear matter effect only.
    \item Calculations with cold and hot medium effects that assume the QGP is described by the hydrodynamic-based model.
\end{itemize}
We will let future experiments falsify either scenario.

\section{Results and discussion}\label{sec:res}
In the result section, we first illustrate the type of modification cold nuclear matter effect and QGP effects may induce in the cross-section ratio $R_{AA}$. We then fix a range for the jet-medium coupling parameter $g_s$ in large colliding systems Au-Au, Pb-Pb, and Xe-Xe. With the same set of parameters, we then present predictions for $R_{AA}^{h}$, $R_{AA}^{D}$, and $R_{AA}^{B}$ in $d$-Au, $p$-Pb, and O-O collisions. These calculations are performed with the dynamical cold nuclear matter effects. We discuss the impact of using the dynamical approach and the nuclear PDF in appendix \ref{sec:res:nPDF}.

\subsection{Interplay of cold (initial-state) and hot (final-state) medium effects}
\begin{figure}
    \centering
    \includegraphics[width=.95\textwidth]{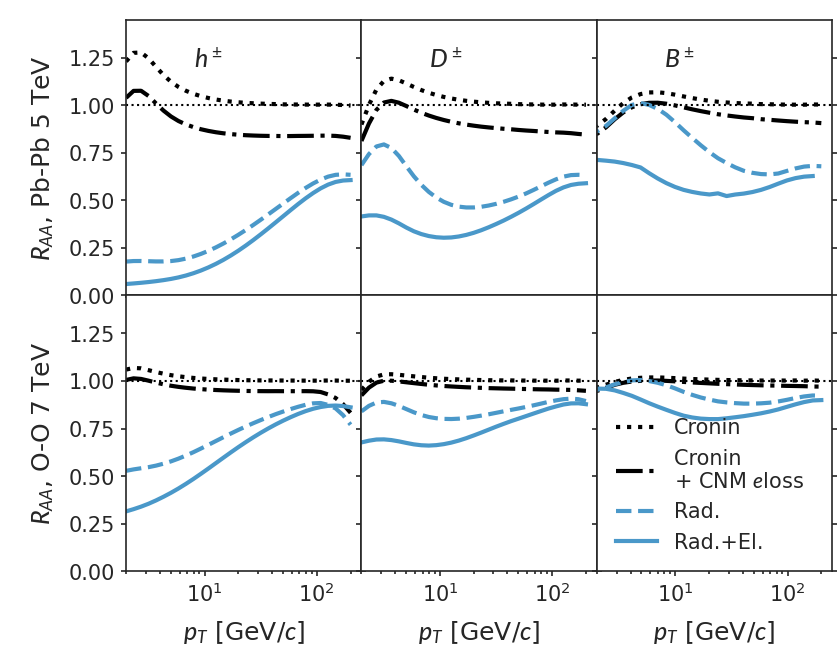}
    \caption{Nuclear modification factor $R_{AA}$ of light hadron (left column), charm  (center column) and bottom (right column) mesons in 0-10\% central Pb+Pb collisions (top row) at 5.02 TeV and O+O collisions (bottom row) at 7 TeV. Within each panel, we show $R_{AA}$ results that sequentially include the Cronin and dynamical shadowing effects (black dotted lines), CNM energy loss (black dash-dotted lines), medium-induced radiation (blue dashed lines), and collisional energy loss (all effects, blue solid lines).}
    \label{fig:Raa-rad-el}
\end{figure}

In figure \ref{fig:Raa-rad-el}, we sequentially include the contribution from Cronin effect and coherent power corrections, CNM energy loss, elastic and radiative effect in the QGP in 0-10\% central Pb-Pb and O-O collisions (rows). The parameter used in this demonstration is $g_s=1.8$. 
Columns from left to right show the modifications for charged hadrons, D mesons, and B mesons.
The dashed lines include only Cronin momentum broadening and the peak for light hardons is around 3 GeV. For heavier mesons, it moves to slightly higher $p_T$.   
The inclusion of CNM energy loss (black dash-dotted lines) results in an overall suppression as the CNM energy loss fraction is almost independent of energy (see equation \ref{eq:CNMeloss}). CNM effects are much smaller in O-O collisions than those in Pb-Pb collisions, as expected from the $A^{1/3}$ scaling.

In Pb-Pb collisions, the modified QCD splitting functions (calculations shown as blue dashed lines) lead to significant suppression of light hadron production. The further inclusion of collisional energy loss (solid blue lines) is a sub-leading effect in $\ln(E)$.
For heavy mesons, the radiative correction in the region $p_T < 5M$ is strongly suppressed, and the modifications are largely attributed to collisions energy loss. One should be careful, however, when interpreting the heavy-flavor results at $p_T\lesssim M$. In this region, the ``jet approximation'' $E\gg M$ completely breaks down, and the heavy quark's orientation can change randomly as it ``diffuses'' in the QGP. The low $p_T$ regime is better modeled by Langevin or other transport approaches that fully evolve the phase-space density of the heavy quark \cite{Rapp:2008qc,Cao:2013ita,Ke:2018tsh}.
In O-O collisions, there is a notable change in the relative importance of radiative correction and collisional energy loss. This can be understood from the different medium-size scaling of the two processes as discussed in section \ref{sec:FS:el-rad}. Especially for heavy meson $R_{AA}$,  collisional processes are responsible for at least 50\% of the QGP modifications in 0-10\% O-O events.

\subsection{Nuclear modifications in Pb-Pb, Xe-Xe, and Au-Au}\label{sec:res:large}
\begin{figure}
    \centering
    \includegraphics[width=1\textwidth]{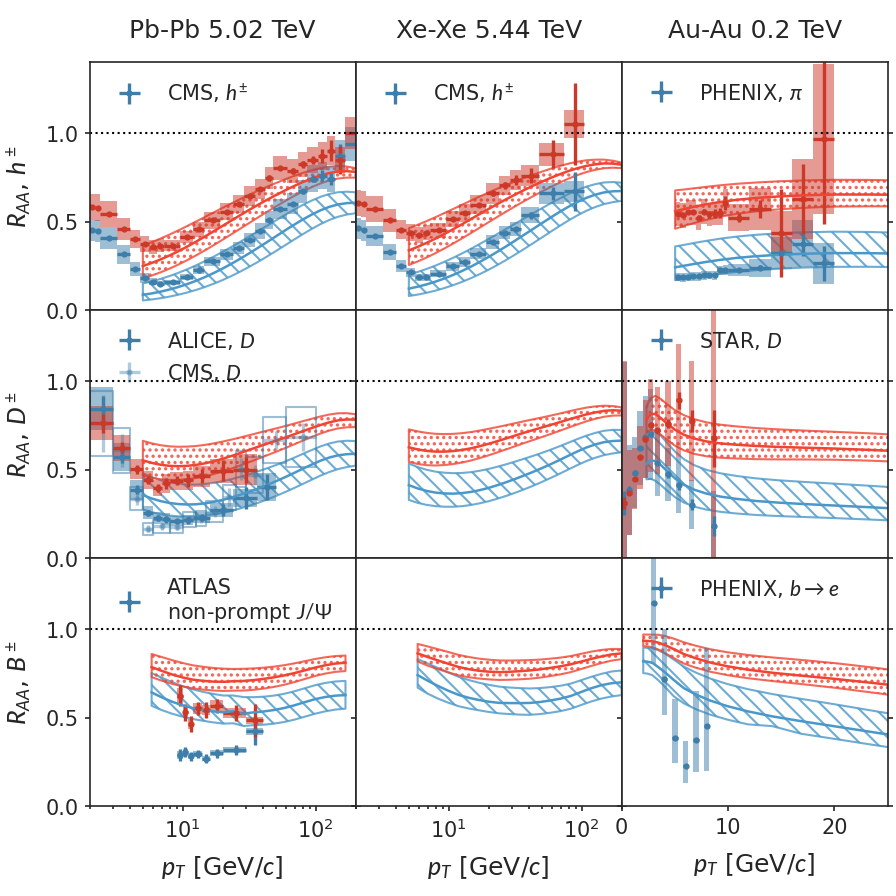}
    \caption{Nuclear modification factor $R_{AA}$ of light hadron (left), charm  (middle) and bottom (right) mesons or their decay products in large colliding systems. Proton and nuclear parton distribution functions are taken from the NNPDF Collaboration. 
    Left column: Pb+Pb collisions at $\sqrt{s}=$5.02 TeV for centrality classes 0-10\% and 30-50\%; middle column: Xe+Xe collisions at $\sqrt{s}=$5.44 TeV for centrality classes 0-10\% and 30-50\%; right column: Au+Au collisions at $\sqrt{s}=$200 GeV for centrality classes 0-10\% and 40-50\%. The calculations are compared to measurements by the ALICE \cite{Acharya:2018hre}, ATLAS \cite{ATLAS:2018hqe}, CMS \cite{CMS:2016xef,Sirunyan:2017xss}, PHENIX \cite{PHENIX:2012jha,PHENIX:2015ynp}, and STAR \cite{STAR:2018zdy} experiments. For bottom flavor, the B-decayed $J/\Psi$ is computed for the LHC energies and B-decayed electron is presented at RHIC energy.
    Note that the centrality classes for the STAR measurements of $D$ meson is 0-10\% and 40-80\%.}
    \label{fig:Raa-L}
\end{figure}
For simplicity of the uncertainty estimation we vary the jet-medium coupling $g_s= 1.8 \pm 0.2$. In principle, $g_s$ may run with the medium scale and this has been investigated in other studies \cite{Ke:2020clc}.  In figure \ref{fig:Raa-L}, calculations with bands that show the sensitivity to the interaction strength are given for $R_{AA}^h$ (top row), $R_{AA}^D$ (middle row), and $R_{AA}^B$ (bottom row) in Pb-Pb collisions at 5.02 TeV (left column), Xe-Xe collisions at 5.44 TeV (middle column), and Au-Au collisions at 200 GeV (right column). 
We include the Cronin effect, cold nuclear matter energy loss, and coherent power corrections. 
Within each panel, the nuclear modification in 0-10\% and 30-50\% central collisions are shown and compared to available data from the ALICE \cite{Acharya:2018hre}, ATLAS \cite{ATLAS:2018hqe}, CMS \cite{CMS:2016xef,Sirunyan:2017xss}, PHENIX~\cite{PHENIX:2012jha,PHENIX:2015ynp},  and STAR Collaboration \cite{STAR:2018zdy}.
For comparisons to $b$-decay electrons and non-prompt $J/\Psi$, smearing functions extracted from Pythia8 \cite{Sjostrand:2014zea} simulations are applied to the $B$ meson spectra. 

For light hadron suppression, the range $1.6<g_s<1.8$ provides a good description of the LHC data in Pb-Pb and Xe-Xe collisions. At RHIC energy the data suggests a larger coupling $1.8<g_s<2.0$. This trend is consistent with many other findings that the effect of jet-medium interactions is larger at lower temperatures relevant for collisions at the RHIC beam energy \cite{PhysRevC.90.014909,Cao:2021keo}. 

Switching to the flavor/mass dependence of the suppression, the calculation agrees well with D-meson suppression at high transverse momentum but slightly overestimates $R_{AA}^{D}$ at low $p_T$. Data tend to lie on the lower edge of the band. The overestimation (not enough suppression) is evident for the bottom quarks. However, we are not going to tune a separate set of parameters for the heavy sector in this paper. Instead, tension with data is a useful indicator of physics that might be missing in the calculation.
One possible explanation for the systematic deviation at low $p_T$ with increasing quark mass is the collisional dissociation of heavy mesons in the QGP~\cite{Adil:2006ra,Sharma:2009hn}. In section~\ref{sec:FS-QGP:DGLAP}, we have argued that after the evolution the light parton fragmentation should take place outside of the QGP medium. However, the formation time of low-$p_T$ heavy mesons is considerably shorter so that they can be produced inside the nuclear medium. The calculation in reference \cite{Adil:2006ra} considers the collisional broadening and break-up of the $D$ and $B$ in the QGP that further suppresses $R_{AA}^D$ below 10 GeV and $R_{AA}^B$ below 30 GeV. This effect is not included in the present study. However, it should be much less important in small colliding systems to be discussed in subsection \ref{sec:res:small}. 
Other works consider collisions between $D$ and $\pi, \rho$ meson in the hadronic phase \cite{Lin:2000jp,Cao:2015hia}, which is important in the low-$p_T$ region.
In addition, reference \cite{Cao:2017crw} studied $M/E$-type drag-induced radiations of heavy quarks. These two additional effects qualitatively push the calculation in the right direction, but their overall magnitudes are too small to explain the large difference that we saw in bottom-flavor $R_{AA}$. 

\subsection{Small systems}
\label{sec:res:small}
\begin{figure}
    \centering
    \includegraphics[width=.9\textwidth]{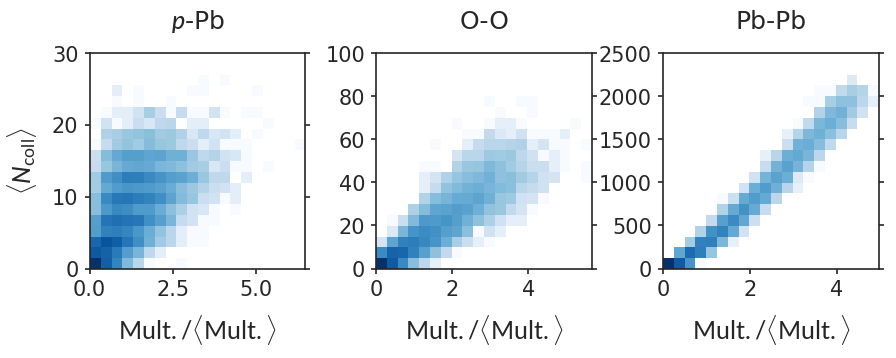}
    \caption{Correlation between TRENTo centrality classes and the averaged number of binary collisions for $p$-Pb, O-O, and Pb-Pb collisions. The O-O collisions are expected to establish a more well-defined centrality selection in small collision systems.}
    \label{fig:Geo}
\end{figure}

With the same range of parameters, we now turn to systematic predictions for small collision systems. At LHC energies there have been extensive measurements of jet production in $p$-$A$ collisions. 
However, the interpretation of the results suffers from the ambiguity of the geometric model of nuclear collisions in the presence of large fluctuation. This situation is illustrated in figure \ref{fig:Geo} obtained using the TRENTo initial condition model used in this study. From the left to the right panel, we plot the histograms of the self-normalized ``multiplicity'' at the initial condition level versus the number of binary collisions ($N_{\rm coll} \equiv T_{AB}/\sigma_{pp}^{\rm inel}$) for $p$-Pb, O-O, and Pb-Pb collisions. 
In large colliding systems, there is a strong correlation between the nuclear geometry and the final-state multiplicity, and the determination of $N_{\rm coll}$ or $T_{AB}$ that normalizes $R_{AA}$ is less sensitive to subnucleonic modeling and fluctuations. This relation strongly decorrelates in $p$-Pb collisions, making the determination of $T_{AB}$ extremely sensitive to proton shape, fluctuations, and particle production mechanisms.
One of the motivations of O-O program is to partly recover the correlation between collision geometry and the multiplicity to provide unambiguous signatures of nuclear modification in small systems.

\begin{figure}
    \centering
    \includegraphics[width=.95\textwidth]{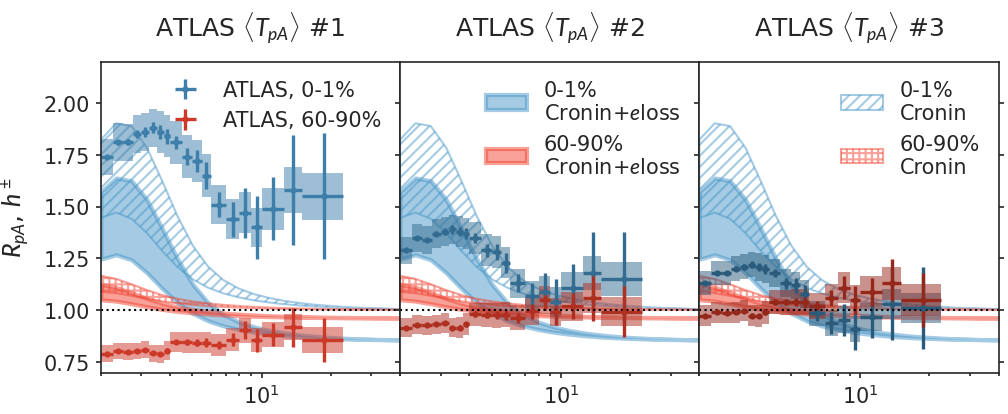}
    \caption{Nuclear modification factor $R_{p\textrm{-Pb}}$ compared to ATLAS data \cite{ATLAS:2016xpn} scaled by the overlap functions from three different calculations of nuclear collision geometry. The results for 0-1\% and 60-90\% centralities are colored in blue and red, respectively. The calculations only include cold nuclear matter effects. The shaded bands include dynamical shadowing and Cronin effect, while the filled bands further reflect consideration of CNM energy loss.}
    \label{fig:Raa-SG-cold}
\end{figure}

\begin{figure}
    \centering
    \includegraphics[width=.95\textwidth]{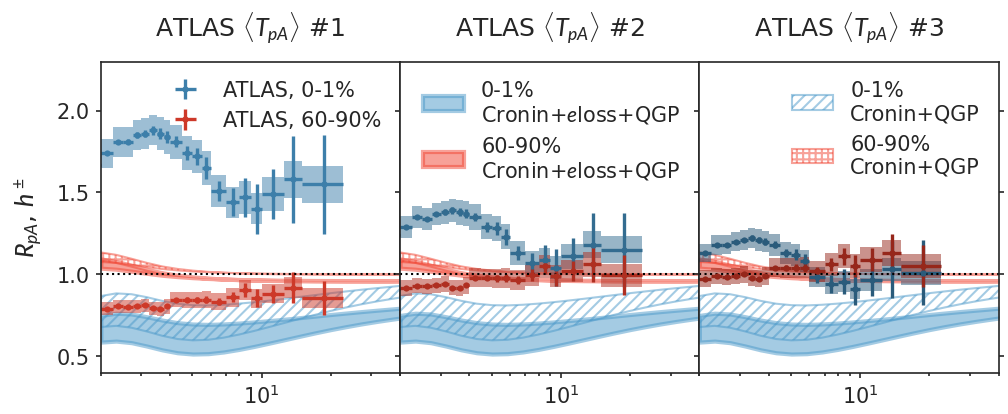}
    \caption{Same as figure \ref{fig:Raa-SG-cold}, but with QGP effect (elastic and radiative).}
    \label{fig:Raa-SG-cold+hot}
\end{figure}

First, we study $R_{pA}$ in $p$-Pb collisions. In figure \ref{fig:Raa-SG-cold}  we compare theoretical predictions with {\it only CNM effects} to the ATLAS data. In theoretical calculations, we always know the correct normalization for $R_{pA}$ such that $R_{pA}=1$ in the absence of nuclear effects.  
On the experimental side, the published $R_{pA}$ data are strongly model-dependent: the ATLAS Collaboration obtains the normalization $\langle T_{\rm pA}\rangle$ in the conventional Glauber model (left), and the improved Glauber-Gribov model with two choices of a parameter that controls the proton fluctuation (middle and right panels) \cite{ATLAS:2016xpn}. 
Here we label them as ATLAS $\langle T_{\rm pA}\rangle$ \#1, \#2, and, \#3. The resulting $Q_{pPb}=dN_{\rm pA\rightarrow h} / \langle T_{\rm pA}\rangle / d\sigma_{\rm pp\rightarrow h}$ is shown to be extremely sensitivity to the experimental choice of the nuclear geometry models. 

This model dependence clearly cannot be controlled within this study. Nevertheless, we argue that it is unlikely that medium corrections can truly cause 50\% enhancement of hadron production at $p_T=20$ GeV as suggested by geometric model \# 1. We consider the geometric models \# 2 and  \# 3 to be much more realistic from the point of view that $R_{AA}\approx 1$ at large $p_T$. 
Focusing on scenarios \#2 and 3, the cold nuclear matter calculation nicely explains the peak at low $p_T$ in the top 1\% high-multiplicity events and its disappearance in 60-90\% $p$-Pb collisions, though small residual enhancement remains in peripheral collisions. Also, the peak in the data is at a slightly higher $p_T$.  This ``centrality'' dependence comes from the nuclear-thickness (impact-parameter) dependence of the Cronin effect. 
Either calculations with or without CNM energy loss could be consistent with the current data in scenarios \#2 and 3, though geometry \#2  favors no CNM energy loss. Clearly, a better understanding of nuclear geometry in $p$-A is needed to further constrain cold nuclear matter effects.

\begin{figure}[h!]
    \centering
    \includegraphics[width=.95\textwidth]{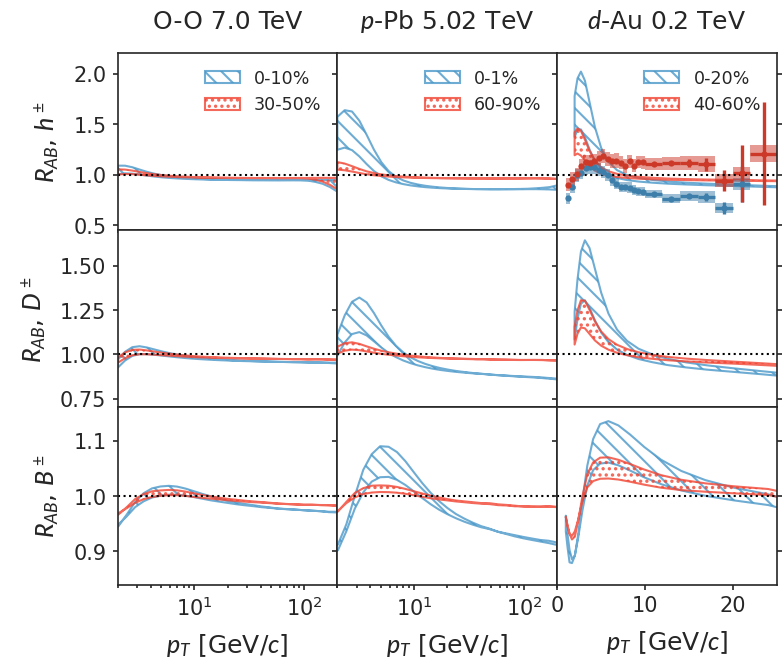}
    \caption{Nuclear modification factor $R_{AB}$ of light hadron (top row), charm  (middle row), and bottom (bottom row) mesons  in small colliding systems.
    Calculations only include cold nuclear matter effects.
    Left column: O-O collisions at $\sqrt{s}=$7 TeV; middle column: $p$-Pb collisions at $\sqrt{s}=$5.02 TeV; right column: $d$-Au collisions at $\sqrt{s}=$200 GeV. The $d$-Au data is obtained by the PHENIX Collaboration \cite{PHENIX:2021dod}.}
    \label{fig:Raa-S-cold}
\end{figure}
\begin{figure}[h!]
    \centering
    \includegraphics[width=.95\textwidth]{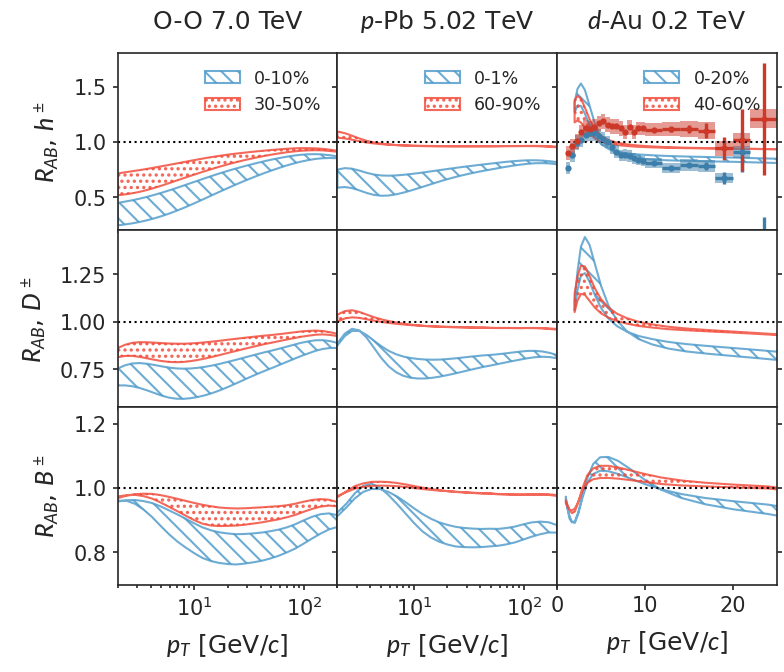}
    \caption{Nuclear modification factor $R_{AB}$ of light hadron (top row), charm  (middle row) and bottom (bottom row) mesons  in small colliding systems.
    Calculations include both cold nuclear matter effects and quenching in the QGP.
    Left column: O-O collisions at $\sqrt{s}=$7 TeV; middle column: $p$-Pb collisions at $\sqrt{s}=$5.02 TeV; right column: $d$-Au collisions at $\sqrt{s}=$200 GeV. The $d$-Au data is obtained by the PHENIX Collaboration \cite{PHENIX:2021dod}.}
    \label{fig:Raa-S-hot}
\end{figure}

Despite the large model-dependent uncertainty, the current measurements in $p$-Pb leave little room for the hot QGP effects discussed in section \ref{sec:med-QGP}. In figure \ref{fig:Raa-SG-cold+hot}, the calculations include both cold and hot nuclear effects. QGP effects introduce a strong centrality dependent suppression of $R_{pPb}$ at intermediate and large $p_T$. This is not consistent with data in either scenario. Again, we emphasize that in our calculation, the temperature of the hot QGP is determined by matching the initial condition to hydrodynamic equations using lattice EoS of state. This cannot exclude models of large non-equilibrium corrections to the density of collision centers in small systems.

Finally, we present our predictions for O-O collisions.
In figure \ref{fig:Raa-S-cold}, we show calculations {\it with only cold nuclear matter effect} for O-O, $p$-Pb and $d$-Au. The magnitude of such CNM effects depends on the transport properties of cold nuclear matter and phenomenology can only be improved with a better understanding of centrality in $p(d)$-A. Nevertheless, 
contrary to $p$-Pb and $d$-Au collisions, the CNM effects are small in O-O collisions. This further establishes that the O-O system is ideal to search for QGP effects.
Results including hot QGP effects are shown in figure \ref{fig:Raa-S-hot}. We estimate that the existence of a deconfined plasma phase in O-O collisions can lead to almost 50\% suppression of the charged-particle $R_{AA}$ at $p_T=10$ GeV in 0-10\% centrality class, while bottom mesons can be suppressed by 20\% at $p_T=20$ GeV.

Both figures \ref{fig:Raa-S-cold} and \ref{fig:Raa-S-hot} are our predictions in small systems with the current understanding of CNM and QGP effects. If there is no QGP
formed in small systems, we expect negligible modifications to $R_{AA}$ in O-O and low $p_T$ enhancement in $p$-Pb and $d$-Au due to the Cronin effect. If QGP
 is created, then the suppression in central O-O collisions can reach 50\% at $p_T=10$ GeV, and there would be a clear ordering to $R_{AA}$ of light, charm, and bottom flavors at intermediate $p_T$.

\section{Summary}\label{sec:summary}
In this paper, we investigated systematically the modification of light and heavy-flavor production in small and large colliding systems at moderate and high $p_T$. Our goal was to differentiate the impact of cold nuclear matter and hot QGP effects. We performed the calculations by including in-medium corrections to the  QCD  factorization framework for the more elementary $p$-$p$ collisions. Cronin effect, coherent power corrections and parton energy loss in the cold nuclear matter modify the initial-state parton densities, while HTL-type collisional energy loss and medium-induced radiative correction change the hadron fragmentation function in the medium. A modified DGLAP evolution approach handles the scale evolution of the fragmentation function in the medium. With jet-medium coupling $g_s = 1.8 \pm 0.2$,  the calculation results in a reasonable agreement with the light flavor suppression in $A$-$A$ collisions at RHIC and LHC. This range can even accommodate the quenching of charm mesons, albeit with the largest of the studied couplings studied. However, the same range of $g_s$ underestimates the bottom meson suppression pointing to remaining tensions with the description of bottom quark dynamics even after the inclusion of collisional energy loss. 

In small colliding systems, we found that the CNM effect only can already explain the basic patterns observed in $p$-Pb collisions scaled by the improved Glauber-Gribov model. Room for improvement in the description of such systems is available as with the CNM transport parameters used here the magnitude of the Cronin enhancement and/or cold nuclear matter energy loss can be overestimated. In order to place better constraints on parton transport in large nuclei, an improved understanding of centrality determination in $p$-A reactions will be greatly beneficial.  In spite of the remaining uncertainties, we managed to establish that the current model of QGP formation in $p$-A as described by hydrodynamics, leads to quenching of hadron spectra that is inconsistent with the $p$-Pb data. The same cannot be said for $d$-Au data, but these two sets of measurements have very different and opposite high-$p_T$ behavior vs centrality. This once again points to the importance of understanding the centrality determination in highly asymmetric small-on-large systems.

As for O-O collisions, we found that the CNM effects alone are only responsible for very small corrections, while the formation of a QGP can suppress charged particle spectra by more than a factor of two and bottom flavor up to 20\%. Unlike the suppression in large systems that is dominated by induced radiation, collisional energy loss in O-O collisions leads to comparable modifications as the effect of medium-induced evolution.  The predicted suppression in small systems at LHC energies with and without QGP formation is very distinct. We finally observed that if QGP quenching effects are identified in O-O, the enhanced contribution from collisional processes can be tested by simultaneously looking at the flavor dependence of $R_{AA}$.

\acknowledgments 
This work is supported by the U.S. Department of Energy, Office of Science, Office of Nuclear Physics through  Contract No. 89233218CNA000001 and by the Laboratory Directed Research and Development Program at LANL.

\appendix
\section{Parameters for hydrodynamic simulations of O-O collisions at 7 TeV}
\label{sec:app:hydro}
\begin{figure}
    \centering
    \includegraphics[width=.6\textwidth]{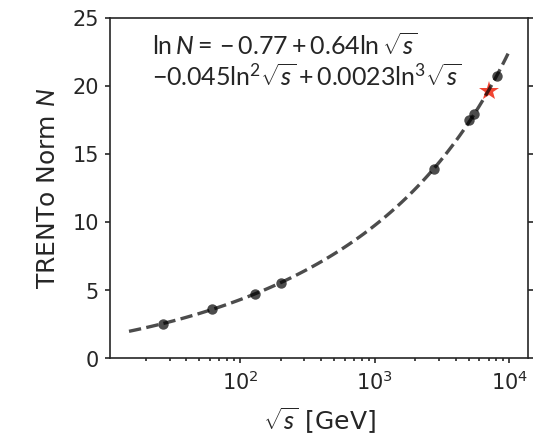}
    \caption{Normalization parameter of TRENTo model tuned to Au-Au and Pb-Pb collisions at RHIC and LHC (black diamond-shaped symbols). They are fitted by $\ln{\rm Norm} = a + b\ln\sqrt{s} + c (\ln\sqrt{s})^2 + d (\ln\sqrt{s})^3$ (dashed line) to extrapolate to the normalization parameter at $\sqrt{s}=7$ TeV (red star).
    }
    \label{fig:TRENTo-norm}
\end{figure}

\begin{figure}
    \centering
    \includegraphics[width=\textwidth]{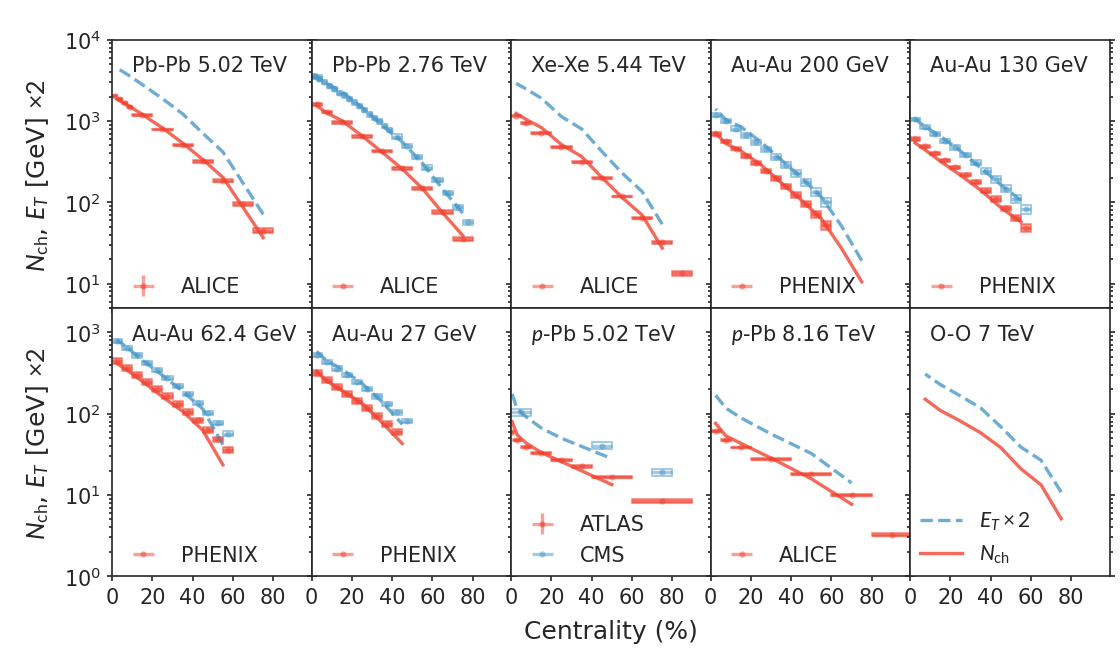}
    \caption{The multiplicity ($N_{\rm ch}$, red solid lines) and transverse energy ($E_T$, blue dashed lines) obtained in hic-eventgen compared to experimental measurements \cite{}. For symmetric systems, the data at midrapidity are shown. For $p$-Pb collisions, we take the data within $\pm 0.5$ units of rapidity around the center-of-mass rapidity.
    The last panel shows the predicted $N_{\rm ch}$ and $E_T(\times 2)$ of O-O at 7 TeV using the interpolated TRENTo normalization.}
    \label{fig:Nch-ET-cen}
\end{figure}

We work in the approximation that the transport parameters are only functions of local temperatures and leave them unchanged from those calibrated in \cite{Bernhard:2018hnz}. We assume that only the normalization parameters change notably at different beam energy and fit them  using a third-degree polynomial in $\ln\sqrt{s}$. The polynomials are constrained by fitting the normalization to the transverse energy ($E_T$) production and charged-particle ($N_{\rm ch}$) multiplicity in Au-Au collisions at 27, 62.4, 130, 200 GeV, Pb-Pb collisions at 2.76 and 5.02 TeV, Xe-Xe collisions at 5.44 TeV, and p-Pb collisions at 5.02 and 8.16 TeV. The simulation is performed with centrality-class-averaged initial conditions, and the quality of the description of $E_T$ and $N_{\rm ch}$ at various beam energies is shown in figure~\ref{fig:Nch-ET-cen}. The polynomial fitting using the extracted normalization factors at different beam energy is shown in figure~\ref{fig:TRENTo-norm}, which results in a normalization at 7 TeV to be 19.6.

The resulting multiplicity and transverse energy as functions of centrality in O-O is shown in the last panel of figure \ref{fig:Nch-ET-cen}. In 0-10\% centrality collisions, $dN_{\rm ch}/d\eta$ is estimated to reach 170. For 30-40\% mid-central collisions, $dN_{\rm ch}/d\eta\approx 58$, which is similar to that in the top 1\% high-multiplicity p-Pb collisions.

\section{Dynamical CNM effect versus nuclear PDF approach}\label{sec:res:nPDF}
\begin{figure}
    \centering
    \includegraphics[width=.9\textwidth]{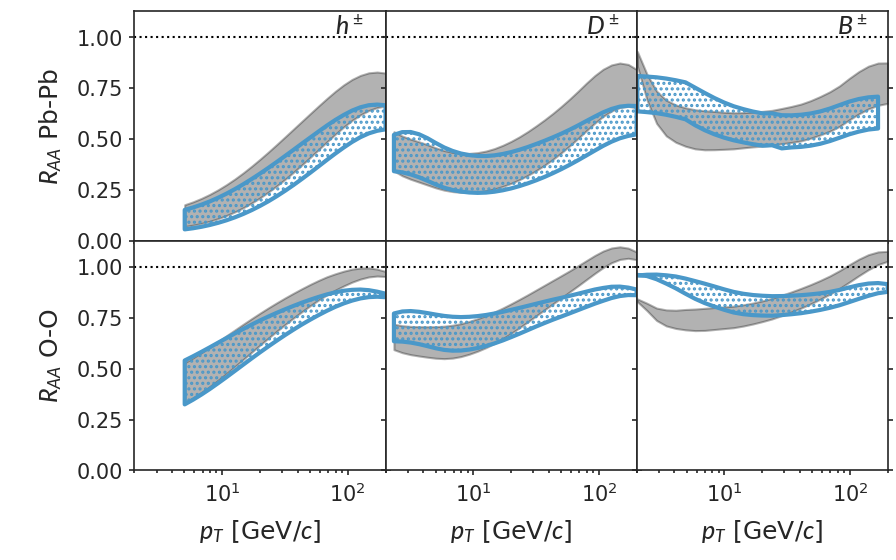}
    \caption{Impact of different cold nuclear matter effect calculation on the nuclear modification factor in the large and small colliding systems. Blue bands used the dynamical CNM model. The read dotted bands applied collinear nuclear PDF from the (n)NNPDF parametrization.}
    \label{fig:Raa-CNM}
\end{figure}

We have done most of our analysis using the dynamical approach for the CNM effects. Finally, we discuss how the signal of hot QGP effects will differ if one performs the calculation with nuclear PDF. In figure \ref{fig:Raa-CNM}, $R_{AA}$ using dynamical CNM approach (blue dotted bands) are compared to nPDF calculation (gray bands) for Pb-Pb and O-O collisions.
For light hadrons and charm mesons, the major differences appear at high $p_T$, since the dynamical approach does not include modifications in the valence region. 
$B$-meson displays a surprising sensitivity to the CNM models at low $p_T$. 
Again, these differences show up in the region where $M/p_T = \mathcal{O}(1)$, which is not a well-controlled region in the current framework. Nevertheless, this difference suggests that when hot medium effects are suppressed by the large mass of $b$ quark, it is possible to be used to probe the details of the CNM calculation. 

\bibliographystyle{JHEP}
\bibliography{ref}
\end{document}